# The degenerate Fermi gas of π electrons in fullerenes and the σ surface instabilities

Shoaib Ahmad
National Centre for Physics, QAU Campus, Shahdara Valley Road, Islamabad 44000, Pakistan
Email: sahmad.ncp@gmail.com

Sabih D Khan
PINSTECH, P O Nilore, Islamabad, Pakistan

Sadia Manzoor
Dept. of Physics, COMSATS, Islamabad, Pakistan

## Abstract

The departure from perfect spherical symmetry in the case of fullerenes ($C_{60}$ being the sole exception) induces instabilities due to the stresses generated by the pentagonal protrusions in the $\sigma$-bonded surfaces. By assuming $\sigma - \pi$ separability and treating $\pi$ electrons as a degenerate Fermi gas in the two shells around the central $\sigma$ structure, the resulting degeneracy pressures can further enhance the $\sigma$-surface initiated instabilities for non-icosahedral structures (especially for those $\langle C_{60}\rangle$) as well as the icosahedral fullerenes ($\rangle C_{60}$) with large protrusions. Under certain circumstances the net degeneracy pressure across the $\sigma$ surface might have structure-stabilizing effect. The role of the $\pi$ electron degeneracy in a broad range of fullerenes from $C_{20}$ to $C_{1500}$ and its effects on fullerene stability is investigated.

## 1. Introduction

By treating the delocalized π electrons of fullerenes as a Fermi gas we investigate the role of the degeneracy pressure $P_\pi$ in determining the stability of fullerenes. This analysis is considered in conjunction with a nanoelastic model [1] that was developed for the nonlinear, pentagon-related stresses on the σ-bonded surfaces of closed carbon cages. In this communication, a broad range of fullerenes from $C_{20}$ to $C_{1500}$ have been considered, however, in the case of fullerenes larger than $C_{60}$ only those structure were chosen that have icosahedral symmetry. On the other hand, six of the smaller than $C_{60}$, non-icosahedral structures have also been discussed. The localized σ- skeleton is treated as continuum elastic structure while the delocalized π electrons in this approximation, appear as the free electron gas in the two shells around the central σ shell. Hückel's original hypothesis [2] on the σ-π separability and orthogonality is assumed. The extension of σ-π orthogonality to 3D closed cages of fullerenes by various authors [3-5] yields analytical solutions for hybridization of the s and p bonds in non-planar conjugated systems. Similarly, tight binding approximation has also been extended by various researchers [6-10] to graphene, fullerenes and the nanotubes. These elaborate and modern extensions of the HMO theory built upon the fabric of linear orbitals LCAO and various combination schemes have provided fairly accurate descriptions of the energies, optical and chemical properties of carbon's closed cages. Thus there have been attempts to extend the HMO model with various additions and modifications to describe the properties of fullerenes with emphasis on $C_{60}$. However, the question of the stability of the entire fullerene family from $C_{20}$ to $C_{1500}$ cannot be answered without using topological arguments like those of the non-abutting pentagons by Kroto [11] to explain the predominance of $C_{60}$ and $C_{70}$ in soot as opposed to other fullerenes that are seen in the mass spectra of the C clusters [12] and that of the non-abutting corannulenes for the stability of 14Å nanotubes capped with 1/2 $C_{240}$ proposed by Ahmad in 2005 [1]. We have taken cues from an alternate approach that was suggested by Pauling [13] and applied by many researchers [14-21] on the conjugated hydrocarbons utilizing the free electron model for the delocalized π electrons of conjugated systems.

Comprehensive treatments of a large number of representative organic molecules using the two alternative techniques of LCAO molecular orbitals and the free electron model [22-24] found considerable agreement between the two approaches. Therefore, treatment of the π electrons in conjugated systems as a Fermi electron gas is not new and has been successfully attempted before [13-24]. As this alternate approach did not provide any breakthroughs in determining the chemical properties of the planar, conjugated, saturated as well as unsaturated hydrocarbons, it has not been the preferred tool of the chemist. However, it will be shown in this communication that the free electron gas of the π electrons in two distinct shells is essential for understanding and explaining the stability of fullerenes when used in conjunction with the nanoelastic model [1] that describes the inherent tendency of the non-spherical fullerenes to buckle. This communication is also based on the same σ-π separability and orthogonality as proposed and applied by various authors [2-22] but with the additional assumptions that:

(1) The localized σ- bonded skeleton of the C nano-shells is expected to show marked differences from its planar counterpart-graphene in its elastic properties that are manifested in the generation of the curvature related stresses. The hallmark of the curved structures is the critical stress $P_\pi^{crit}$ that has a non-linear dependence on ζ- the orthogonal protrusions of the pentagonal defects.

(2) The treatment of the delocalized π electrons as a degenerate Fermi gas can provide degeneracy pressures from the inner and outer shells around the σ- skeleton that can either stabilizes the 3D network of fullerenes or induce buckling.

(3) The total number of π electrons ($N_\pi$) may be distributed in approximate ratios of the respective volumes of these two shells (the inner and the outer shells) and the associated π- electron numbers may not be equal i.e., $N_\pi^{in} \neq N_\pi^{out}$ in fullerene shells. This ensures equal densities in the inner and outer shells.

(4) The thickness of the σ-shell $t_\sigma$ is assumed to be ~C core ($1s^2$) diameter [25] being of the order of 0.2 Å and that of the π-shells $t_\pi$~1/2 diameter of a C atom minus $t_\sigma$ (~ 0.85 Å); therefore, the volumes satisfy $V_\pi \gg V_\sigma$ for all fullerene shells.

## 2. Stability of fullerenes: C$_{20}$ to C$_{540}$ and beyond

Out of the 28 fullerene point groups the icosahedral ones with $I_h$ and $I$ groups have the highest symmetry [5]. C$_{20}$, the smallest fullerene, is a dodecahedron with 12 all-abutting pentagons in full icosahedral symmetry ($I_h$). The higher fullerenes have an increasing number of hexagons with the twelve omnipresent pentagons. These belong to a wide range of symmetry groups while the ones belonging to the group $I_h$ (the full icosahedral group of the order 120, i.e. the number of symmetry axes), are C$_{60}$, C$_{240}$, C$_{540}$, C$_{960}$....besides C$_{20}$. The symmetry argument is being used to explore the stability of fullerenes as a function of the sphericity. This line of argument has been used earlier [1] to explain the elastic properties of fullerenes as well as to highlight the role that the pentagonal protrusions play in $>$ C$_{60}$ fullerenes. The icosahedrally symmetric protrusions are illustrated in Figure 1 where three fullerenes C$_{20}$, C$_{60}$ and C$_{540}$ are shown.

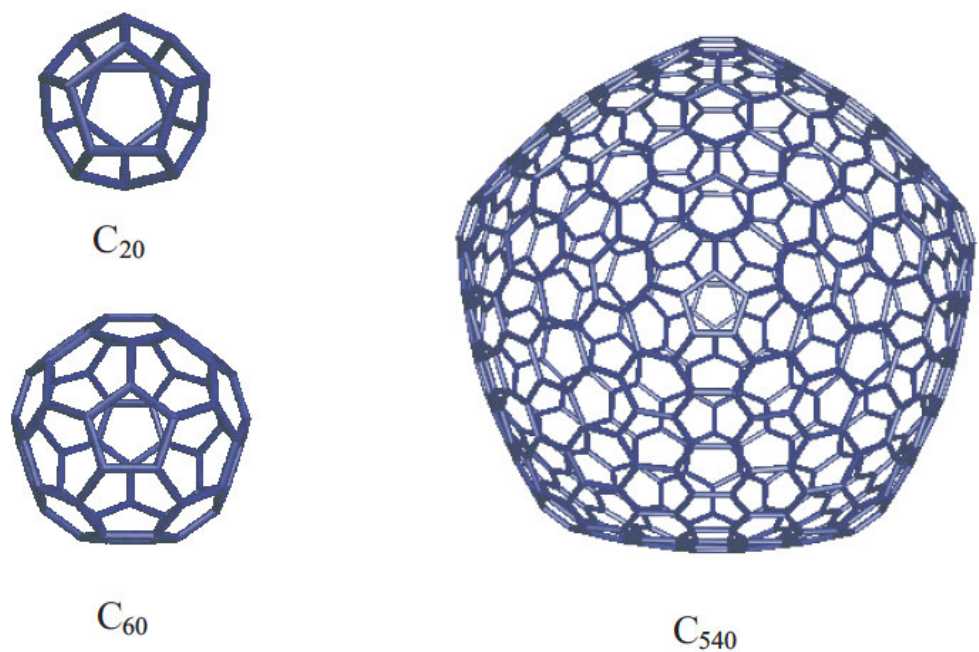

**Figure 1.** Three significant members of the fullerene family C$_{20}$, C$_{60}$ and C$_{540}$ are shown. All three belong to the same icosahedral symmetry group $I_h$ but vary largely in terms of the number of C atoms, radii, sphericity and elastic properties.

C$_{60}$ being the perfectly spherical structure, C$_{20}$ has 20 protrusions along its icosahedral dual's edges [5] while C$_{540}$ clearly demonstrates the 12 pentagonally protruding regions with the protrusion $\zeta$ to radius ratio 1:4. This ratio goes on increasing for the higher fullerenes with the consequently increased instability.

Using these arguments it was predicted by Ahmad in 2002 [1] that the larger free standing fullerenes may emerge from sooting discharges with inherent instabilities.

The above arguments about the stability of fullerenes are based on the continuum elastic and topological properties that can be deduced for the σ - surfaces. In this model the π electrons were included in the overall thickness $t(\equiv t_\sigma + t_\pi)$ of the shells since that is the only way to incorporate their structure-related parameter i.e. the flexural rigidity $D = Yt^3/12(1 - v^2)$; where *Y* is Young's modulus and *v* the Poisson ratio for graphene. The role of the delocalized π electrons in determining the stability of these structures is the next stage of investigation that is being probed in this paper. These π electrons from now onwards will be treated as a degenerate Fermi gas in the present paper.

## 3. The nanoelastic model of fullerenes

The internal stresses which occur when a body is deformed locally are due to the forces of interaction between atoms. These have short ranges of the order of inter-atomic distances. Such forces when acting on any part of the body are considered to act locally on the surface. In the theory of elasticity, being a macroscopic theory, generally the distances considered are large compared with inter-atomic separations. However, in fullerenes, by considering these local deformation-induced internal stresses, insight into their elastic stability related properties can be achieved. The analogy of the nanometre scaled C shells with macroscopic hollow structures and shells has prompted various authors [1, 26-32] to use continuum elasticity theory. Nanoelasticity of fullerenes may be considered an attempt to relate the structure related deformations of the closed cages of carbon with the resulting surface stresses. Majority of these authors have used the continuum elasticity theory to describe fullerenes as bent graphene. We also have extended the continuum mechanics to describe various growth related properties of fullerenes and nanotubes by introducing the stretching and bending effects of the spherical structures' deformations [1]. Elastic properties of spherical shells including fullerenes and nanotubes are different from those of graphene, which may be considered equivalent to thin plates or flat sheets.

Theory of continuum elasticity [33] treats the deformation of plates in two distinct regimes; (a) thin plates with small deformations $\zeta \leq t$ where *t* is the plate thickness, and (b) when large deformations ($\zeta \gg t$) occur due to the application of large external forces or as we will show, in the case of fullerenes, due to the spherical curvature induced by pentagons. Pure bending is the dominant effect in the former case while both bending and stretching effects are important in the case of the latter. It is shown in this communication that by using continuum elasticity of macro-surfaces, plates and shells we can understand the elastic behaviour of the spherically bent sp$^2$−bonded C sheets and spheroidal fullerenes with pentagonal protrusions.

In the case of small deformations, ζ denotes the vertical displacement of a point on the neutral surface on which neither extension nor compression takes place. On a neutral surface the displacement vector **u** has components $u_x = u_y \approx 0$ and $u_z = \zeta(x, y)$. In such cases the strain tensor is $u_{\alpha\beta} = \frac{1}{2}(\frac{\partial u_\alpha}{\partial x_\beta} + \frac{\partial u_\beta}{\partial x_\alpha})$. The total free energy of a plate for small deformations as a function of $\zeta$, that is the displacement of points on the surface of the plate, is

$$E_{plate} = \tfrac{1}{2} D \iint \left[ \left\{ \frac{\partial \zeta^2}{\partial x^2} + \frac{\partial \zeta^2}{\partial y^2} \right\}^2 + 2(1-\nu)\left\{ \left(\frac{\partial^2 \zeta}{\partial x \partial y}\right)^2 - \frac{\partial^2 \zeta}{\partial x^2}\frac{\partial^2 \zeta}{\partial y^2} \right\} \right] dxdy \qquad 1)$$

where $D$ is the flexural rigidity of the plate $D = Yt^3/12(1-\nu^2)$. By minimizing this energy one obtains the equation of equilibrium of the plate. Dividing the integral in two parts and varying these separately we get

$$f = D\Delta^2 \zeta \qquad (2),$$

where the 2D Laplacian $\Delta \equiv \frac{\partial^2}{\partial x^2} + \frac{\partial^2}{\partial y^2}$. This is the equation for equilibrium of plate bent by the application of the external force on it. This equation is solved by setting up appropriate boundary conditions that can be very complex. Considerable simplification is achieved if the plate edges are clamped or supported. In the case of the supported plate, for example, shown in Figure 2(c), one has $\Delta^2 \zeta = 0$ everywhere except at the origin and the result for the force in terms of $\zeta$ and plate dimensions is

$$f \approx 16\pi D \zeta \left/ \left[ (R^2 - r^2) - 2r^2 \log \frac{R^2}{r^2} \right] \right. . \qquad (3)$$

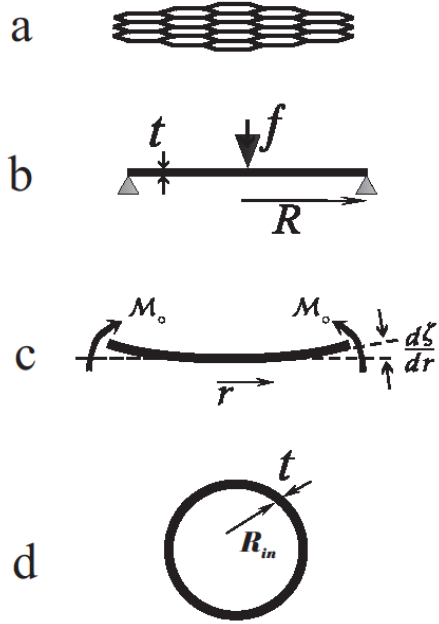

**Figure 2.** The sequence of deformations is shown leading to small deformation in a 19 hexagon graphene sheet of thickness $t$ and flexural rigidity $D = Yt^3/12(1-\nu^2)$, where $Y$ is Young's modulus and $\nu$ the Poisson ratio. In Fig. 2(a) an external applied force $f$ is acting to bend it with the associated bending moments $M_o \approx (1+\nu)D/r$ at its edges. For circular supports as shown in Fig. 2(c) spherical deformation of curvature $1/r = DM_o/(1 + \nu)$ will occur. If the supports shown in Fig. 2(b) are parallel then cylindrical bending of the sheet into an open ended tubule with inner radius $R_{in}$ could result in Fig. 2(d).

Figure 2 shows the sequence of deformations that can lead to small spherical deformation in a graphene sheet with 19 hexagons, of thickness *t* in Fig. 2(a) under an external applied force *f* in Fig. 2(b) to a plate under bending moments $M_o \approx (1 + v)D/r$ at its edges as shown in Fig. 2(c). If the supports shown in Fig. 2(b) are not circular and instead are parallel then cylindrical bending of the sheet into an open ended tubule with inner radius $R_{in}$ could result in Fig. 2(d). The order of magnitude estimate of pure bending energy $E_{plate}$ in Eq. (1) can simplify the physical considerations. The first derivatives of $\zeta$ are of the order of $\zeta/R$ and the second derivatives $\sim \zeta/R^2$. Thus $E_{plate} \sim Yt^3\zeta/R^2$; this is the order of magnitude estimate for the bending energy per unit area of flat plates under external force $f$ per unit area. It shows a linear relationship of the bending energy with the deformation.

## 3.1 Large deformations of graphene sheets

In this section we will first establish nanoelastic limits to large deformations of graphene sheets (plates of nanometer thickness) to yield corannulene-like structures that may lead to the formation of fullerenes or capped nanotubes, and then compare and consider the respective energies of such bent graphene with the spherically curved bowl of corannulene due to a central pentagon. Since both bending and stretching are involved and there is no neutral surfaces in large deflections unlike the situation where $\zeta \leq t$. The infinitesimal strain tensor for the simultaneously bent and stretched plate is expressed as a function of displacement vector **u** for pure stretching and transverse displacement $\zeta$. The 2D infinitesimal strain tensor is

$u_{\alpha\beta} = \frac{1}{2}(\frac{\partial u_\alpha}{\partial x_\beta} + \frac{\partial u_\beta}{\partial x_\alpha}) + \frac{1}{2}\frac{\partial \zeta}{\partial x_\alpha}\frac{\partial \zeta}{\partial x_\beta}$. The first term is similar to the strain tensor for small deformations except that it also includes second-order terms for the derivatives of $\zeta$.

We evaluate the two equations that govern the equilibrium of a graphene sheet of thickness t under an external force f per unit area with transverse deformation $\zeta (\gg t)$,

$$f = D\Delta^2\zeta - t\frac{\partial}{\partial x_\beta}\left(\sigma_{\alpha\beta}\frac{\partial \zeta}{\partial x_\alpha}\right) \text{ and } \partial\sigma_{\alpha\beta}/\partial x_\beta = 0 \tag{4}$$

where $\sigma_{\alpha\beta}$ is the stress tensor.

Introducing the stress function $\chi$ defined by $\sigma_{xx} = \partial^2\chi/\partial y^2$, $\sigma_{xy} = -\partial^2\chi/\partial x\,\partial y$, $\sigma_{yy} = \partial^2\chi/\partial x^2$ one eventually obtains $\Delta\chi = \sigma_{xx} + \sigma_{yy}$ and

$$f = D\Delta^2\zeta - t\left(\frac{\partial^2\chi}{\partial y^2}\frac{\partial^2\zeta}{\partial x^2} + \frac{\partial^2\chi}{\partial x^2}\frac{\partial^2\zeta}{\partial y^2} - 2\frac{\partial^2\chi}{\partial x\,\partial y}\frac{\partial^2\zeta}{\partial x\,\partial y}\right) \tag{5}$$

To obtain the equation of equilibrium in terms of the stress function and the deformation we have

$$\Delta^2\chi + Y\left(\left[\frac{\partial^2\zeta}{\partial x^2}\frac{\partial^2\zeta}{\partial y^2}\right] - \left[\frac{\partial^2\zeta}{\partial x\,\partial y}\right]^2\right) = 0 \tag{6}$$

Equations (5) and (6) form a complete set of equations for large deflections of plates. The set is, however, very complex and cannot be solved exactly. These are non-linear equations and even for simple systems one has to make approximations to get order of magnitude estimates for the energies, deformations etc. This is the starting point for our deliberations on such systems and we will evaluate for given deformations the estimates of the forces that may be responsible. The first estimate is for the situation when dealing with large deformations $\zeta \gg t$, we will determine the relation between the deformation and the associated force. Estimation of the terms in Eq. (6) shows that $\chi \sim Y \zeta^2$, similarly the first term in Eq.(5) is smaller than the second which is of the order of magnitude $Yt\zeta^3/R^4$, where $R$ is the radius of the plate. Since this is comparable with the external force, we have

$$f \sim Y t\zeta^3/R^4. \tag{7}$$

This is a non-linear relation where force is proportional to the cube of $\zeta$. Comparing equation (7) with (3) immediately shows that there is a linear relationship between $f$ and $\zeta$ for small deformations while it is not the case for large deformations. In addition, we get only approximate solutions in the case of large deformations.

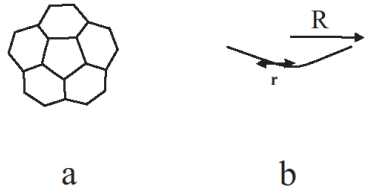

**Figure 3.** (a) and (b) show the top and side views of a corannulene bowl with 20 C atoms. The figure emphasizes the two important dimensions. The curvature can be viewed as the protrusion centred around a pentagon with $r$~1 – 2Å in the larger spherically curved sheet of R~ 3 – 4Å consisting of hexagons only.

## 3.2 The role of spherical curvature in corannulenes and fullerenes

Initiation of curvature in any embryonic C structure due to the formation of a pentagon is the essential starting point for the cage closure resulting in the formation of spherical caps for nanotubes or fullerenes. In figure 3(a) and 3(b) the top and side views of a corannulene are shown with the two important dimensions; the curvature can be viewed as a spherical protrusion with $r$ ~ 1 − 2Å in the larger circular curved sheet of R~ 3 – 4Å consisting of hexagons only. For fullerenes with icosahedral symmetry $Ih$ the centre of the pentagon has C5 symmetry. We have recently discussed [1] the role of symmetry on the formation of either the fullerenes or single-walled nanotubes from the embryonic corannulene-like or coronene-like structures. The figure illustrates the curvature-related stresses and relates the magnitude of protrusion and deformation induced by external forces as $r \approx \zeta$.

A typical fullerene ($>C_{60}$) is shown in figure 4(a) with protrusions $\zeta \geq t$, superimposed on a sphere of radius $R$, thickness $t$. These protrusions can be produced when the spherical shell is subjected to a concentrated force per unit area $f_o$ along the inward (Fig. 4(b)) or outward normal (Fig. 4(c)). The resulting deformation $\zeta = R_0 - R$, where $R$ and $R_0$ are the radii of the inscribing and circumscribing spheres of the respective Goldberg polyhedra. In the case of fullerenes with protruding corannulenes like $C_{80}$, $C_{140}$, $C_{180}$ etc, one can determine the size of the pentagonal protrusion $\zeta$ from topographical features and relate these to the stresses that are generated. These internal stresses are of the order of the force $f_o$ of Fig. 4(b) and 4(c). A major part of the elastic energy is stored in the narrow bending strip $\sim d$ on the edge of the bulge. Geometrically, angle $\alpha$ is the angle subtended by the bulge at the spherical centre and $= \zeta/d \approx r/R$; area of this strip $\sim rd$.

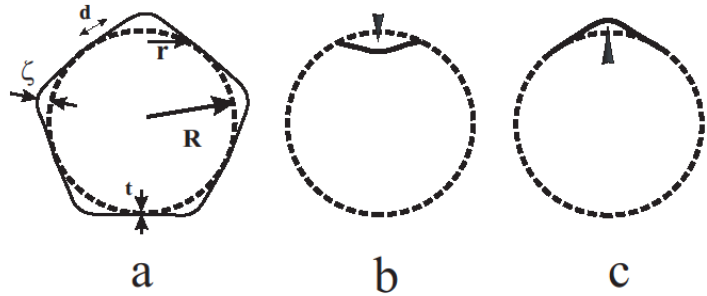

**Figure 4.** A typical, large ($>C_{60}$), icosahedral fullerene with twelve protrusions is shown in figure 4(a) with protrusions $\zeta \geq t$, superimposed on a sphere of radius R, thickness t. In Fig. 4(b) and 4(c) is shown that these protrusions arise by subjecting a perfect spherical shell to a concentrated force $f_o$ along the inward (Fig. 4(b)) or outward (Fig. 4(c)) normal.

Deformation $\zeta$ will vary from zero to maximum over distance $d$. The inscribing shell's radius $R$ can be considered to be equivalent to the radius of the corresponding fullerene. The distance $d$ is a measure of the region of the protrusion and the area of deformation $\sim d^2$. The pure bending energy over this area varies as $E_{ben} \sim Yt^3 (\zeta/d)^2$ as discussed earlier. The stretching accompanying the bending in flat plates is a second order effect as evident from Eq. (1) but in the case of shells the order is reversed and stretching becomes the first order effect. The infinitesimal strain tensor $\sim \zeta/R$ and the corresponding stress tensor $\sim Y\zeta/R$ and the deformation energy per unit area $\sim Yt(\zeta/R)^2$. Over an area $d^2$ the total stretching energy $E_{str} \sim Yt[(\zeta d)/R]^2$. The bending energy decreases and stretching increases with the increase in $d$ thus both energies are considered in determining the deformation near the point of application of the force. Minimizing their sum $E_{ben}+E_{str}$, one gets $d \sim \sqrt{tR}$. As can be visualized from Fig. 4, bending is along the meridian and stretching along the circle of latitude (of radius $r$). The total elastic energy in the bending strip of a corannulene is $E_{cor} \approx Y t^{5/2}(\zeta^{3/2}/R)$. For a given value of the deformation one can obtain the outward force $f_o$ by equating it to the derivative of $E_{cor}$ with respect to $\zeta$ as

$$f_o \approx Y t^{5/2}(\zeta^{1/2}/R). \tag{8}$$

As the bulge or the outward protrusion has occurred due to the generation of internal stresses associated with the curvature of the corannulene, this stress $P$ can be associated with the work done in producing a defect volume $\sim \Delta V$, where $\Delta V \sim r^2\zeta \sim \zeta^2 R$. The total free energy $= E_{cor} - P\zeta^2 R$. Derivative of this total free energy yields the critical stress $P_{crit}$,

$$P_{crit} \approx Y t^{5/2}/(R^2 \zeta^{1/2}). \tag{9}$$

This is an inverse relation between $\zeta$ and $P$ (i.e., $\zeta$ increases when $P$ decreases) hence it indicates an unstable equilibrium; therefore, the bulges with large $\zeta$ grow on their own accord, while the smaller ones shrink. Equation (9) corresponds to a maximum of the total free energy. The emergence of a bulging curved surface can be related with an internal, outward force $f_o$ whose magnitude is obtained by taking the derivative of $E_{cor}$ with respect to $\zeta$. The nanoelastic model that was developed using the above arguments provides a picture of a typical soot forming environment where C accretion produces planar as well as curved C structures. It is the curvature related surface forces that play a decisive role in determining whether fullerenes or nanotubes will be formed and of what dimensions [1]. However, in this communication, our major concern is to investigate and discern the roles of the $\sigma$ and $\pi$ shells with regard to a broad range of fullerenes and their impact on the stability of these spherical structures.

## 4. The curvature-related properties of the π shells

The energetics and structural stability of fullerenes has been the subject of extensive studies since the discovery of fullerenes by Kroto et al. in 1985 [12]. Topological arguments [1,34-39] have provided justifications for the spherical structures to bee more stable that their planar counterparts. Such arguments are centred on the experimentally observed stability of C60. Delocalization, rehybridization and interactions of the $\pi$ electrons with the spherical $\sigma$ networks have been studied to provide explanations of the sphericity of the shelled structures. Haddon [40] investigated the role of the strain induced by curvature in C60, C70 and their organometallic derivatives by using his $\pi$-orbital axis vector analysis that treats the interactions of $\pi$ and $\sigma$ electrons in the formation of their 3D, non-planar orbitals. The strain energy of $C_{60}$ is shown to be between 77 to 100% of the total heat of formation [29,40]. Therefore, curvature is a seen as the source of strain in fullerenes with spheroidal structures. But much less effort has been spent on the strain introduced by the structural protrusions in fullerenes (with respective isomers) that are $<$ or $>$ $C_{60}$. The present paper is an effort to bridge the gap between the deformations induced by the pentagonal protrusions in icosahedral fullerenes and the non-icosahedral ones. Curvature is the outcome of the bonding in freely growing C sheets with $>$20 C atoms by the addition of monomers ($C_1$s), diatoms ($C_2$s) and tri-atomic ($C_3$s) molecules in carbonaceous vapour that results pentagon formation in addition to that of hexagons. The process is similar to the bending of graphene sheets with large and permanent deformations with the difference that some pentagons or heptagons are also introduced along with the hexagons. To visualize and see the growth of a fullerene and the associated curvature we have simulated the formation of $C_{60}$ by the addition of C atoms in ones, twos and threes to interactively learn about the way the $\sigma$ and $\pi$ surfaces evolve as the pentagons and hexagons are added to chains of C atoms. For the purposes of this communication, we show in figure 5 the results of an MNDO simulation depicted as the snap shots of three crucial stages of fullerene formation by the addition of $C_2$s.

The structures shown in figure 5 have been obtained by geometry optimization calculations in ArgusLab 3.1 [41] using semi-empirical method named modified neglect of diatomic overlap (MNDO) [42]. The $\sigma$ bonds are illustrated with solid sticks and the overlapping cloud is the electron density obtained by using energy calculations in MNDO. The picture of the growth of $C_{60}$ presented here is for the purpose of illustration of the curvature related properties of the $\pi$ shells. It takes place via an open-caged route by $C_2$ insertion and is one of the ways cage closure may occur in carbon vapour. For the purpose of illustration of the role of the curvature induced stresses, we have chosen to show the open cage sequence of events that leads to $C_{60}$'s cage with only the three snap shots shown in figure 5. A detailed and thorough investigation is presented by Khan and Ahmad elsewhere [43].

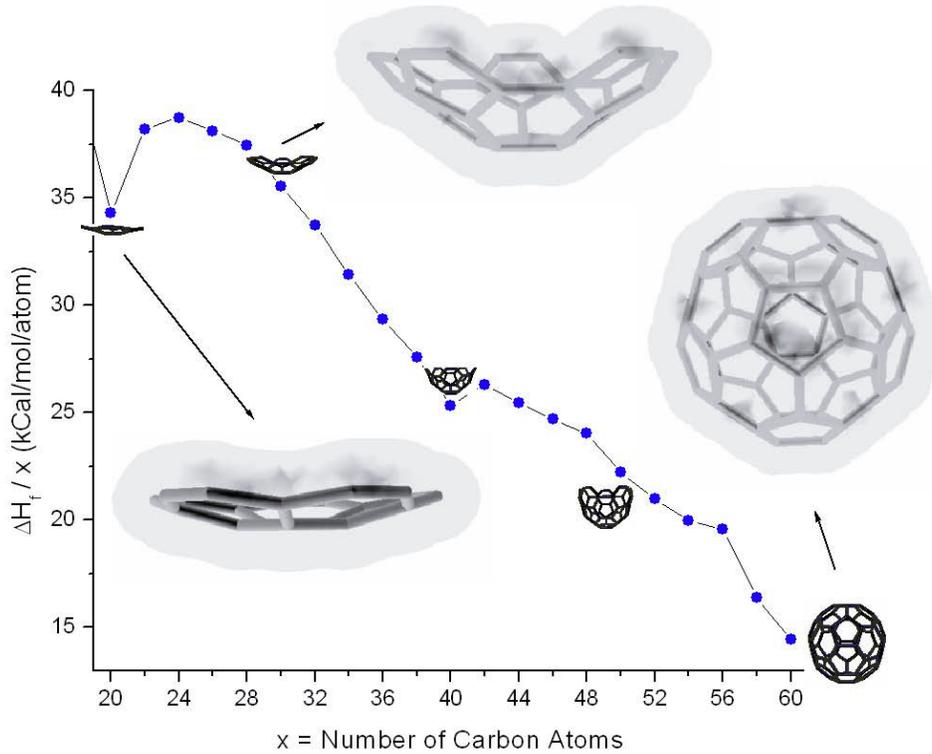

**Figure 5.** The separation of σ and π shells in three important formative stages of formation of $C_{60}$ are shown. The top and side views of corannulene-like bowls consisting of 20 C atoms are shown in Fig. 5(a) and 5(b). Addition of further 10 C atoms leads to a half-cap in Fig. 5(c) and 5(d) consisting of 30 C atoms, 6 pentagons. This highly strained structure can lead to the formation of a $C_{60}$ shell shown in Fig. 5(e).

The first snap shot is that of a corannulene that has just been formed; figure 5(a) and 5(b) are the top and side views of corannulene with the central $\sigma$ surface surrounded by the $\pi$ network. This is the first time when the $\sigma$ structure of the flat growing sheet is strained by the curvature. The top and side views of the next important stage are depicted in figure 5(c) and 5(d) which is equivalent to the half cap of $C_{60}$ with five pentagons and five hexagons consisting of 30 C atoms. At this stage the structure is under intense surface generated forces and associated bending moments, described by Ahmad in 2005 [1]. Figure 5(e) is the final stage of the completion of the fullerene structure where the two $\pi$ shells surround the central $\sigma$ structure consisting of 12 pentagons and 20 hexagons. The figure shows clearly that as the graphene sheet starts to curl the associated inner and outer $\pi$ sheets have different volumes especially, near the final stages of cage closure. This effect will be more pronounced for smaller fullerenes. In the next section we present a quantum mechanical description of the $\pi$ electrons treated as a degenerate Fermi gas and discuss the two cases of the distribution of $\pi$ electrons in equal numbers i.e. $N_\pi^{in} = N_\pi^{out} = N/2$, and approximately equal densities $n_\pi^{in} \approx n_\pi^{out}$ in the two shells of fullerenes from $C_{20}$ to $C_{1500}$.

## 5. The delocalized π electrons as a degenerate Fermi gas

The interaction-free electron gas is the simplest approximation that neglects all interactions, the Coulomb interactions of the electrons with each other and the interaction of the electrons with the positive background of the ionic lattice. Yet this model of free electron gas explains many phenomena and yield physically meaningful parameters of solids like the density of states, Fermi energies and the associated degeneracy pressures [44]. Similar free- electron models have been used to describe the properties of conjugated hydrocarbons based on the high mobility or delocalization of the π electrons [13-24]. The chemical properties explained by using such a simple model include calculations of the π electron wavefunctions, energy levels, dipole moments etc. In addition, by including a suitable perturbing potential, one can extend the simple interaction-free electron gas to take care of the reactivity and similar chemical properties of π electron systems. In Schmidt's model [14] an aromatic system is regarded as a box filled with Fermi gas of a finite number of π electrons. We extend this model to the delocalized π electrons of fullerenes to learn about their role in the stability of the respective structures. An extra merit of the free-electron approximation is that one can obtain continuous electron distributions for different fullerene structures.

For a typical fullerene shown in figure 6(a) the σ- bonded spherical skeleton with radius rσ is surrounded by the inner and outer π electron shells of radii $r_{in}$ and $r_{out}$. The ratio of the outer to inner π shell volumes for the fullerenes from $C_{20}$ to $C_{1500}$ as a function of the number of C atoms in respective fullerenes is shown in figure 6(b) to be $V_\pi^{out}/V_\pi^{in}$ ~2.78 for C20 and asymptotically approaches 1 for the larger ones. The resulting π electron density is also a smoothly varying parameter and the set of two π electron clouds provides a unique system where the properties of the Fermi gas with discrete number of free electrons can be evaluated [44]. The Schrödinger equation for the degenerate π-electron gas can be considered with a Hamiltonian with kinetic energy operator alone, all other interactions being neglected and the wave function contains only the spatial part ϕ(r)

$$-\hbar^2/2m \nabla^2 \phi(r) = E\phi(r) \qquad (10)$$

E being the one-electron energy; has a plane wave solution $\phi(r) = e^{ik.r}$ with $E = \frac{\hbar^2 k^2}{2m}$. To normalize ϕ(r) the electron gas is restricted to a cube of volume $V_\pi$, yielding the normalized wave function $\phi(r) = \frac{1}{\sqrt{Vg}} e^{ik.r}$. Components of momentum $k_\alpha = \frac{2\pi}{L_\alpha}$, α = x, y, z. For perfect degeneracy the fermions or the π electrons in our case, occupy all states up to a limiting momentum $k_\pi$.

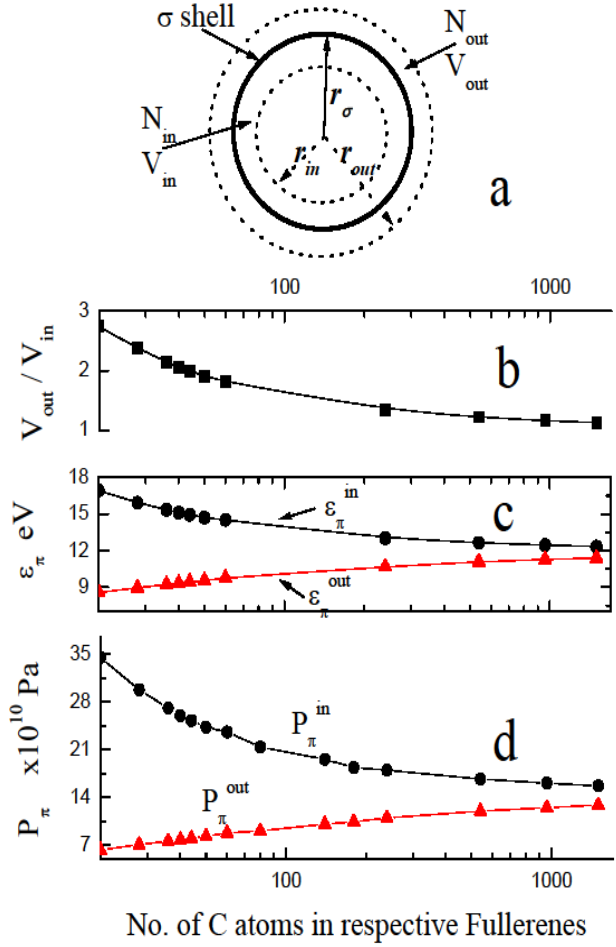

**Figure 6.** In figure 6(a) a typical fullerene is shown with its σ– bonded spherical skeleton of radius rσ that is surrounded by the inner and outer π electron shells of radii $r_{in}$ and $r_{out}$. The number and volumes of inner and outer π electron shells are also indicated. Figure 6(b) shows $V_\pi^{out}/V_\pi^{in}$ as a function of the number of C atoms in respective fullerenes from $C_{20}$ to $C_{1500}$. In figure 6(c) the Fermi energy $\varepsilon_\pi$ for the inner and outer shells is shown for the same fullerene range. Figure 6(d) shows $P_\pi$ for the two respective π shells.

The number of quantum states of translational motion of π electrons in the interval k to k +dk are $4\pi k^2 dk V_\pi/(2\pi\hbar)^3$. The total number of π electrons in respective shells and the associated Fermi momenta from zero to $k_F$ is

$$N_\pi = \frac{V_\pi}{2\pi^2\hbar^3}\int_0^{k_F} k^2\, dk = \frac{V_\pi k_F^3}{6\pi^2\hbar^3}. \tag{11}$$

The Fermi momentum $k_F$ is given in terms of the electron density $(N_\pi/V_\pi)$ by $k_F = (\frac{6\pi^2}{2})^{1/3}(\frac{N^\pi}{V^\pi})^{1/3}\hbar$ and the corresponding Fermi energy

$$\varepsilon_F = \frac{k_F^2}{2m} = (3\pi^2)^{2/3}\frac{\hbar^2}{2m}\left(\frac{N_\pi}{V_\pi}\right)^{2/3}. \tag{12}$$

The total energy of the degenerate π electron gas is obtained by multiplying the number of states $4\pi k^2 dk V_\pi/(2\pi\hbar)^3$ by $k^2/2m$ and integrating from zero to $k_F$

$$E_\pi = \frac{V_\pi}{4m\pi^2\hbar^3}\int_0^{k_F} k^4\,dk = \frac{V_\pi k_F^5}{20m\pi^2\hbar^3} = \frac{3}{10}(3\pi^2)^{\frac{2}{3}}\frac{\hbar^2}{m}\left(\frac{N_\pi}{V_\pi}\right)^{2/3} N_\pi. \qquad (13)$$

From the equation of state, the Fermi gas obeys $P_\pi V_\pi = \frac{2}{3}E$, thus the degeneracy pressure of a finite number of π electrons in fullerene shells is

$$P_\pi = \frac{1}{5}(3\pi^2)^{2/3}\frac{\hbar^2}{2m}\left(\frac{N_\pi}{V_\pi}\right)^{5/3}. \qquad (14)$$

The degeneracy pressure of a Fermi gas of π electrons is proportional to the 5/3 power of the electron density. In our case, this will be densities of the respective shells i.e. inner and outer ones $n_\pi^{in}$ and $n_\pi^{out}$.

## 5.1 The case for an equal number of π electrons in the two shells i.e. $N_\pi^{in} = N_\pi^{out}$

If we assume that the available π electrons are distributed equally in the two shells i.e. $N_\pi^{in} = N_\pi^{out} = N/2$ leading to $n_\pi^{in} = N_\pi^{in}/V_\pi^{in}$ and $n_\pi^{out} = N_\pi^{out}/V_\pi^{out}$, then the ratio of respective electron densities $n_\pi^{in}/n_\pi^{out}$ will be in the ratio $V_\pi^{out}/V_\pi^{in}$. Since $\varepsilon_\pi \propto n_\pi^{2/3}$ and $P_\pi \propto n_\pi^{5/3}$ from equations (12) and (14), large differences appear between the Fermi energies and the degeneracy pressures of the two shells for all fullerenes. The differences are, however, most significant for the smaller ones and reduce somewhat for the larger fullerenes. $\varepsilon_\pi$ and $P_\pi$ are plotted from equations (12) and (14) for the two respective π shells of fullerenes from $C_{20}$ to $C_{1500}$ in figure 6(c) and 6(d). Fig. 6(c) shows that $\Delta\varepsilon_\pi (= \varepsilon_\pi^{in} - \varepsilon_\pi^{out})$ varies from 8.42 eV for $C_{20}$ to 4.75 eV for $C_{60}$, and remains high even for $C_{540}$ where $\Delta\varepsilon_\pi \approx 1.6$ eV. Similarly, in Fig. 6(d) the ratio $P_\pi^{in}/P_\pi^{out} \sim 5.5$ for $C_{20}$ and reduces to about 3 for $C_{60}$. These two degeneracy pressures $P_\pi^{in}$ and $P_\pi^{out}$, for the larger fullerenes reduce slowly towards a common value yet the difference remains significant.

Large differences in $\varepsilon_\pi$ and $P_\pi$ for fullerenes will make these structures inherently unstable. Application of the March model to Fullerenes, as done by Clougherty and Zhu [39], predicts ≈ 44% of electrons are in the inner region, with ≈ 56% of electrons in the outer region, for a typical fullerene of radius r = $6.7a_0 \approx 3.56$ Å.

## 5.2 The case for equal charge density in the two π shells i.e. $n_\pi^{in} \cong n_\pi^{out}$

In figure 6 the case of approximately equal charge densities in the two π shells i.e. $n_\pi^{in} \cong n_\pi^{out}$ appears indirectly when large fullerenes are considered for which $V_\pi^{out}/V_\pi^{in}$ approaches 1. This leads to the situation resulting in the merging of the Fermi energies $\varepsilon_\pi^{in} \cong$ $\varepsilon_\pi^{out} \cong 12 eV$ and $P_\pi^{in} \cong P_\pi^{out} \cong 1.4 \times 10^{11}$ Pa. Equal Fermi energies induce stability as the electron transitions across the σ surface are not encouraged and equal degeneracy pressures imply a uniformly squeezed structure. The equality of $\varepsilon_\pi$ and $P_\pi$ for the two shells on its own is not a guarantee for stability; together with the σ surface generated forces the net outcome in the form of stable or unstable fullerenes is discussed below.

## 6. The combined effect of $P_{crit}$ and $\Delta P_\pi$

The case of an equal number of π electrons in the two shells i.e. $N_\pi^{in} = N_\pi^{out}$ introduces massive destabilizing surface forces. Even $C_{60}$ will have a tremendous net outward pressure and would be unstable, which is contrary to experimental observations. The non-sphericity of the abutting pentagons introduces strains in fullerenes from $C_{20}$ to $C_{58}$ and $P_\pi$ remains very large for this group of smaller fullerenes. Therefore, fullerenes $< C_{60}$ are prone to mechanical failure due to both of the surface forces; one that is generated due to the stresses of the σ surface and second due to a large degeneracy pressure difference of the π electrons $\Delta P_\pi$ across the σ surface. These together will not only be the source of instability in smaller fullerenes but also forbid these to be formed at all, which is against the experimental observations. This interpretation of our analysis given above, agrees with the March Model [39], which predicts an increasing discrepancy between $N_\pi^{in}$ and $N_\pi^{out}$ as r increases. It is suggested that the case of approximately equal charge densities in the two πshells i.e. $n_\pi^{in} \cong n_\pi^{out}$ is physically more meaningful and can explain the observed stability of perfect spheroidal cages like $C_{60}$ and provide answers for the instabilities of fullerenes larger or smaller than $C_{60}$.

Figure 7 shows the two types of forces acting on the σ-bonded surface; $\Delta P\pi$ is the net difference between the degeneracy pressures of the inner and outer π shells ($\Delta P_\pi = P_\pi^{in} - P_\pi^{out}$) from Eq. (14) and $P_{crit}$ being the curvature-generated surface stresses from Eq. (9). Whereas $\Delta P_\pi$ is evaluated for the entire fullerene range from $C_{20}$ to $C_{1500}$, the calculations of $P_{crit}$ are done only for the icosahedral fullerenes with *I* and *I*h symmetry.

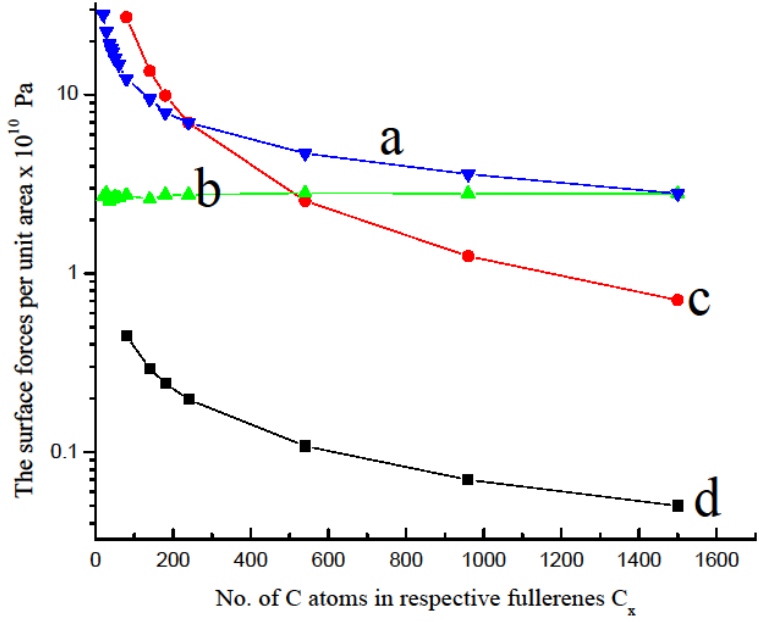

**Figure 7**. The two types of surface forces acting on the σ-bonded structure for the fullerenes from $C_{20}$ to $C_{1500}$ are shown. In figure 7(a) $\Delta P_\pi$ - the net difference between the degeneracy pressures of the inner and outer π shells from Eq. (14), is shown; this is the case for an equal number of π electrons in the two shells i.e. $N_\pi^{in} = N_\pi^{out}$ as discussed in section 5.1. Figure 7(b) has the data for the case of approximately equal charge densities in the two π shells i.e. $n_\pi^{in} = n_\pi^{out}$. The net degeneracy pressure is $\cong 2P_\pi$. Figure 7(c) is the critical σ surface stress $P_{crit}$ calculated using Eq. (9) with the thickness of the entire shell being considered i.e. $t_\sigma = 1.82$ Å. In figure 7(d) we have evaluated $P_{crit}$ for $t_\sigma \approx 0.2$ Å as discussed in section 1.

Figure 7(a) shows $\Delta P_\pi$ calculated under the assumption of equal number of π electrons in the two shells i.e. $N_\pi^{in} = N_\pi^{out}$ while figure 7(b) shows the degeneracy pressure on the σ surface by the oppositely directed pressures $P_\pi \approx 2P_\pi^{in} \approx 2P_\pi^{out}$ where approximately equal charge densities in the two π shells i.e. $n_\pi^{in} = n_\pi^{out}$ have been assumed, therefore the two pressures equalize. Figure 7(c) is the critical σ surface stress $P_{crit}$ calculated with the thickness of the entire shell being considered i.e., $t_\sigma \approx 1.82$Å. This is the same result that was presented in our earlier paper [1]. In figure 7(d) we have evaluated $P_{crit}$ for $t_\sigma \approx 0.2$Å for the reasons discussed in section 1. $\Delta P_\pi$ and $P_\pi$ shown in figure 7(a) and 7(b), respectively, are for fullerenes starting from $C_{20}$, but $P_{crit}$ is shown only for the higher fullerenes starting from $C_{80}$ for the reason that our nanoelasticity model relates the symmetrically disposed deformations with $P_{crit}$, thus only the icosahedral ones and $> C_{60}$ fullerenes are candidates for such an analysis. $P_{crit}$ for $C_{60}$ is not shown as the pentagonal protrusions cannot be worked out in the perfectly spherical fullerene cage.

For smaller fullerenes the magnitudes of $\Delta P_\pi$ and $P_{crit} \sim 10^{11}$ Pa in figures 7(a) and 7(c). These are the estimates with two extreme assumptions; one that distributes the πelectrons equally in the two shells and the other where the σ – π separability is not maintained. In the latter case, the overall effect of all the electrons in the single, unified C shell is represented in the structural rigidity D ∝ t³, where thickness t(≡ $t_\sigma + t_\pi$). In figure 7(b) $\Delta P_\pi$ is ~ $3 \times 10^{10}$ Pa indicating equal but substantial forces acting on the σ-surface due to the oppositely directed pressures of the degenerate Fermi gas of π electrons.

The trends shown in Figure 7 clearly indicate that the assumption of an equal number of π electrons in the two shells i.e. $N_\pi^{in} = N_\pi^{out}$ is obviously not tenable. The tremendous degeneracy pressure difference will simply tear the structure apart. The large Fermi energy differences will encourage π electrons to move out from the inner shell so that the charge density in the two π shells equalizes and $n_\pi^{in} = n_\pi^{out}$.

When does the transition from $N_\pi^{in} = N_\pi^{out}$ to $n_\pi^{in} = n_\pi^{out}$ occur? It is the next crucial stage in cage closure. A tentative guess would be that during the cage closure a competition between the σ surface induced curvature and the differences between the inner and the outer degeneracy pressures ΔPπ makes the π electron densities equalize. If the π electrons were to distribute themselves in the two respective shells of closed cages in such a way that the electron densities tend to equalize i.e. $n_\pi^{in} \approx n_\pi^{out}$ then $\Delta\varepsilon_\pi$ will approach zero and $\Delta P_\pi \approx P_\pi^{in} \approx P_\pi^{out}$.

However, it may be pointed out that even in the case of $n_\pi^{in} \approx n_\pi^{out}$ there is still a tensile force (from the inner π shell) and a compressive (from the outer π shell) one acting on the entire σ surface with $P_\pi^{in} \approx P_\pi^{out}$. Such redistribution will minimize the difference of the two Fermi energies Δεπ for the respective structures. It can be seen that the net effect of approximately equal densities $n_\pi^{in} \approx n_\pi^{out}$ is to reduce the destabilizing forces in the entire range of fullerenes irrespective of the size and symmetry. Such a range of fullerenes has been seen in TOF mass spectra [12].

The case of larger fullerenes (≥$C_{240}$) needs further comment because the assumption of a thin shell the thickness $t_\sigma$ becomes ≪ the radius of larger fullerenes. This may have a stabilizing effect due to the reduction of pentagonal protrusion as a result of large compressive surface forces on both sides of the central σ surface $\Delta P_\pi$ (= $P_\pi^{in} - P_\pi^{out}$) that tend to restore sphericity. But in such cases $\Delta P_\pi$ approaches the value for a graphene sheet. Although the degeneracy pressure differences may not be sufficient to compensate for the structural deformations in the free standing fullerenes, they might be responsible for the sphericity of the composite fullerene structures called the carbon onion shells, designated as $C_{60}@C_{240}@C_{540}@C_{960}@C_{1500}@\cdots$. The high resolution TEM pictures do not reveal the expected pentagonal protrusions along C5 axis [45-47]. On the other hand, for structures (≤ $C_{240}$) the net effect of large $\Delta P_\pi$ may be to increase the instabilities related to structural deformations, especially for non-icosahedral fullerenes. $C_{60}$ is the sole exception having the perfect spherical shape.

# 7. Conclusions

By treating the π electrons of spherically curved graphene sheets and the resulting fullerenes as a degenerate Fermi gas in the two shells around the central σ skeleton, we have visualized the various stages of the evolution of curved C surfaces and fullerenes. We have also shown that the instabilities that are inherent in the protrusion-ridden fullerenes (with the exception of $C_{60}$), can be quantified and their effect on the growth of fullerenes estimated. If the πelectron shells are treated as hollow spherical volumes containing the Fermi gas then their degeneracy pressures Pinπ and Poutπ have profound effects on the growth and stabilities of carbon's curved surfaces in different ways:

(1) During the initial phase of the bowl-like σ-structure formation, the instability associated with the pentagonal protrusions (due to $P_{crit}$) will be further enhanced due to the large outward directed surface forces associated with $\Delta P_\pi$. These structures are predicted to be spinning [1] under the effect of the bending moments induced due to the curvature of the growing C bowls.

(2) As the embryonic structure grows and the cage closes, the protrusions of the σsurface will become preferentially stressed regions under the influence of $P_{crit}$. For the closed cages we have predicted above that the differences between the inner and outer shells' degeneracy pressures reduces along with the differences of densities ($n_\pi^{in} - n_\pi^{out}$). In such cases, the oppositely directed $P_\pi$ becomes the determining factor for the structural instability for all non-icosahedral cages.

(3) The question of the relative stability of various fullerenes with these two types of forces is answered using the data presented in figure 7(b) and 7(d) that shows that the critical stress $P_{crit}$ needed for the onset of instability in fullerenes > $C_{60}$ is at least an order of magnitude smaller than the corresponding $P_\pi$.

.